\documentclass[11pt]{article}
\usepackage{mathrsfs}
\usepackage{amsmath}

\global\arraycolsep=1pt \oddsidemargin.20in \evensidemargin.5in
\usepackage[Symbol]{upgreek}

\usepackage{bbm}
\usepackage{dsfont}
\usepackage{amssymb}
\usepackage{textcomp}

\newcommand{\CR}{\nonumber \\*}

\def\L{{\cal L}}

\DeclareMathAlphabet{\mathpzc}{OT1}{pzc}{m}{it}
\def\s{\,\mathpzc{s}\,}
\def\s{s}

\def\d{\mathfrak{d}}

\def\L{{\cal L}}

\def\d{\delta}

\def\z{\zeta}

\def\be{\begin{equation}}
\def\ee{\end{equation}}
\def\bea{\begin{eqnarray}}
\def\eea{\end{eqnarray}}
\def\bdis{\begin{displaymath}}
\def\edis{\end{displaymath}}
\def\corr{$\clubsuit$}
\def\nn{\nonumber}

\begin{document}
\allowdisplaybreaks[1]
\renewcommand{\thefootnote}{\fnsymbol{footnote}}
\def\corr{$\spadesuit $}
\def\trefle{ $\clubsuit$}
\renewcommand{\thefootnote}{\arabic{footnote}}
\setcounter{footnote}{0}
 \def\stop{$\blacksquare$}
\begin{titlepage}
\null
\begin{center}
{{\Large \bf
  Twisted $N=1$, $d=4$ supergravity and its   symmetries  
}}
\lineskip.75em \vskip 3em \normalsize {\large Laurent
Baulieu$^{ \dagger\ddagger}$\footnote{email address:
baulieu@lpthe.jussieu.fr},
Marc  Bellon$^\ddagger$\footnote{email address:
bellon@lpthe.jussieu.fr}
and
Valentin Reys$^{\ddagger\star}$\footnote{email address:
vreys@nikhef.nl}
\\
\vskip 1em
  $^{\dagger}${\it Theoretical Division CERN }\footnote{ CH-1211
  Gen\`eve, 23,
  Switzerland }
\\
$^{\ddagger}${\it LPTHE
Universit\'e Pierre et Marie Curie }\footnote{ 4
place Jussieu,
F-75252 Paris
Cedex 05, France}
\\
$^{\star}${\it NIKHEF Theory Group }\footnote{ Science Park 105, 1098 XG Amsterdam, The Netherlands}
 }

\vskip 1 em
\end{center}
\vskip 1 em
\begin{abstract}
We display the construction of a twisted superalgebra for the $N=1 $ Euclidean supergravity  on $4$-manifolds with an almost  complex structure. It acts on a representation of twisted supersymmetry made of   forms with odd and even statistics and it is covariant under a $U(2)\subset SO(4)$ Lorentz invariance of the manifold's  tangent-space. It contains 4  twisted supersymmetry generators, one  nilpotent scalar, one  vector and one pseudo-scalar. The superalgebra closes  on the twisted fields of supergravity in its new minimal set of auxiliary fields.  
Its couplings to the twisted Wess and Zumino and vector multiplets are also determined.
\end{abstract}

\end{titlepage}

\def\ba{{\bar a}}
\def\z{\bar z}
\def\bm{{\bar m}}
\def\bn{{\bar n}}
\def\bp{{\bar p}}
\def\bq{{\bar q}}
\def\br{{\bar r}}
\def\bs{{\bar s}}
\def\demi{\textstyle{\frac{1}{2}}}
\def\s{\delta}
\def\sos{\delta_{\rm o.s}}
\def\sos{\delta_{\rm 1.5}}
\def\o{\omega}
\def\la{\lambda}
\def\gg{\chi}
\def\sg{\hat{s}}

\section{Introduction}

Twisting is an important tool in the study of supersymmetric theories and has given important new insights in those studies. It fundamentally means that one supercharge is singled out and used as the primary symmetry of the theory. The twist often  allows for a splitting in the set of supersymmetric generators, which can be very useful. One can find in some cases a subset of the generators that is sufficient to constrain the Lagrangian to be invariant under the full supersymmetry, while it admits off-shell closed field representations.
 
The first examples have used non-trivial \(R\)-symmetries associated to extended supersymmetries to retain a full Lorentz invariance. However it has proved useful to consider the twist of \(N=1\) theories, even if it means that only part of the Lorentz symmetry is explicitly realized, a \(Spin(7)\) or \(U(4)\) symmetry in dimension eight, a \(G_2\) symmetry in dimension seven. 

Here we consider the case of the simplest four-dimensional supergravity, to illustrate the formalism of twisted symmetry in curved space. We work with an Euclidean signature, which allows us to retain a \(U(2)\) subgroup of the rotational symmetry. This same twist has been previously considered in the theory with only global supersymmetry~\cite{lbN}.
 
In the case of the $N=1$, $d=4$ Euclidean supergravity, only a subset of the rotational symmetry is explicitly realized  after the twist, and spinors are no longer present in the theory. All fields transform as tensorial products of the fundamental representation of $U(2)\subset SO(4)$. The fermionic part of the symmetry algebra consists in four fermionic twisted generators, one scalar, one vector and one pseudo-scalar. The translations are part of the supersymmetry algebra and appear, in the twisted formalism, in the anticommutator of the vector supersymmetry generators and the scalar or pseudo-scalar generators. The twisted generators can be untwisted to recover the spinorial anticommuting generators of Poincar\'e supergravity.

The twisted superalgebra and  the superalgebra  of  Poincar\'e supergravity are related in the fact that they define the same invariant action, modulo a twist.  Twisted and untwisted supergravity transformation laws can be related by a linear mapping, in a way that generalizes the case  of super-Yang--Mills theories~\cite{lbN}. 
 
The construction of the twisted superalgebra is done on a $4$-manifold with an Euclidean signature and an almost  complex structure. In this case, the Majorana spinors can be decomposed in holomorphic and antiholomorphic forms.

Among the twisted fermionic generators, the scalar nilpotent one is of main interest to us. It is formally similar to a BRST operator, and has an  analogous  interpretation as the twisted supersymmetry generator of    topological Yang--Mills symmetry, 2d-quantum gravity or topological string~\cite{bausin}. 
The supergravity action in a twisted form is in fact determined by the invariance under this scalar supersymmetry, with   an interesting decomposition occurring for both the Einstein and Rarita--Schwinger actions. Subtle phenomena arise when one requires the additional invariance under the  full $SO(4)$ symmetry group.

Building the   twisted superalgebra  produces a new interesting framework. First, we mention that supersymmetric invariants exist as non-trivial local cocycles, a property that might be of significant importance if the twisted construction  can be extended  to supergravities of rank $N \geq 2 $. Second, the fact that the invariance under the twisted scalar supersymmetry generator alone is enough to write down an action for the twisted fields might be of interest to  bypass the issues raised by the lack of a system of auxiliary fields  in theories such as  higher dimensional supergravities.
It could be that requiring the off-shell closure of the complete Poincar\'e superalgebra is just too demanding. Within this approach, the super-Poincar\'e symmetry is not postulated, but is an emergent property once the invariance under the twisted scalar supersymmetry is imposed.
  
In view of these hypothetical higher dimensional generalizations, we have  computed the twisted formulation for the couplings of supergravity to scalar and vector multiplets. The results are less aesthetic than those obtained for the genuine supergravity multiplet, but their existence is a plausible four-dimensional signal that twisted formulations could also be obtained in $2n\geq4$ dimensions, with a corresponding $U(2)\to U(n)$ generalization.

The scheme of the paper is as follows. In section~2 we recall some known facts about $N = 1$, $d = 4$ supergravity in the new minimal scheme, focusing on the BRST formulation of its symmetries. In section~3, we display a possible (anti)selfdual decomposition of the supergravity action by exploring some properties of the Einstein and Rarita--Schwinger Lagrangians. In sections 4 and~5, the twisted formalism is introduced through definitions of the twisted fields  and the twisted operators corresponding to the symmetries of the supergravity action. The various curvatures needed to build the supergravity action are also displayed in twisted form. 
In section~6,  we use the so-called 1.5 order formalism to build the twisted scalar symmetry generator for all fields but the spin-connection and give a primitive twisted  form of  the supergravity action.  In section~7, we   explore the consequences of requiring the invariance of the  action under the twisted vector symmetry, which eventually yields the complete twisted supergravity action.    In section~8,  we compute   the  coupling to twisted  supergravity of the  twisted    Wess--Zumino and vector   multiplets.  Finally, appendices give useful formulas.
  
\section{$N=1$, $d=4$ supergravity in the new minimal scheme}

The  $N=1,d=4 $ supergravity multiplet  in the  new minimal system of auxiliary fields~\cite{akvoso,gaogso,sostwe} is
\begin{equation}
\label{multiplet}
e^a, \lambda, \omega^{ab}, A, B_2
\end{equation}
Here $e^a$ is the 1-form  vielbein,    the Majorana spinor   $\la = \la_\mu dx^\mu$ is the 1-form  gravitino and $\omega^{ab}$ is the spin-connection 1-form. $A$ and $B_2$  are auxiliary fields, with gauge invariances, such that the multiplet  has as many bosonic and fermionic degrees of freedom both  on-shell and off-shell, modulo the gauge invariances. The abelian 1-form gauge field $A\sim A+dc$ gauges chirality and    $B_2\sim B_2+d \Lambda_1, \Lambda_1 \sim \Lambda_1  +d \Lambda_0$ is  a gauge real 2-form.

The associated curvatures are
\begin{eqnarray}
\label{curvatures}
R^{ab} &=& d\o^{ab} + \tfrac{1}{2}\left[\o,\o\right]^{ab}\CR
T^a &=& de^a + \o^{ab}e_b + \tfrac{i}{2}\bar{\la}\gamma^a\la \CR
\rho &=& d\la + \left(\tfrac{1}{2}\o^{ab} \gamma_{ab} +A\gamma^5\right)\la \\
G_3 &=& d B_2 + \tfrac{i}{2}\bar \lambda \gamma^a \lambda e_a \CR
F &=&dA \nn
\end{eqnarray}  
We will often use the covariant derivative notation $D\equiv d+\o+A$. We use the following expression of the $N=1$ supergravity action, as in \cite{baubel}
\begin{equation}
\label{sugra}
I=\int_{{\cal M}_4}\left(\frac{1}{4}\epsilon_{abcd}e^a{\scriptstyle \wedge}\;e^b{\scriptstyle \wedge}\;R^{cd}(\o) + i\bar{\la}\,{\scriptstyle \wedge}\gamma^5\gamma^a \rho(\la,\o,A){\scriptstyle \wedge}\,e_a - 2B_2{\scriptstyle \wedge}\,dA + {}^*G_3{\scriptstyle \wedge}\,G_3 \right)
\end{equation}

The multiplet~(\ref{multiplet})  is an  off-shell balanced multiplet with 6 bosonic degrees of freedom defined modulo all gauge invariances, 12 fermionic ones, and 6 auxiliary ones, according to the following count: 
\begin{eqnarray} 
e^a:  & &6=16-6  \  { \rm  Lorentz}- 4 \  { \rm   reparametrizations}\CR \la:  & &  12=16-4 \ { \rm    supersymmetries}\CR  A:  & & 3=4-1 \ { \rm   chiral}   \CR B_2: & & 3=6-4\  { \rm   vector  }+1  \ { \rm  scalar} \nonumber  
\end{eqnarray} 
The spin-connection is not an independent field, but is fixed by the (super)covariant constraint 
\begin{equation}
T^a(e,\la)= -\tfrac{1}{2}G^a_{bc}e^b e^c
\end{equation} 
so that $\o^{ab} = \o^{ab}(e,\la,B_2) \equiv \o^{ab}(e,\la) + \tfrac{1}{2}G^{ab}_{c }e^c$, where $\o^{ab}(e,\lambda)$ is the usual spin-connection seen as a function of the vielbein and gravitino. This necessary constraint expresses the fact that no first-order formalism exists for getting an off-shell closed Poincar\'e supersymmetry and an invariant action. 
  
The transformation laws of the various fields under supersymmetry can be expressed using a BRST symmetry operator $s$, where one replaces  all parameters of  supergravity infinitesimal transformations  by  local ghost  fields with opposite statistics. All ghosts transform under the BRST symmetry, in such a way that  $s$ is nilpotent. The nilpotence of \(s\) is equivalent to the off-shell closure of the system of supergravity  infinitesimal transformations, as shown in~\cite{baubel}. This BRST symmetry can be built directly (both in the minimal and new minimal set of auxiliary fields), as outlined below. 
  
Call $\xi^\mu$ the vector ghost for reparametrization.   The other ghosts are those of local SUSY ($\chi$), Lorentz symmetry ($\Omega$), the chiral $U(1)$ symmetry ($c$)  and the 2-form gauge symmetry ($B^1_1$).
The $\xi^\mu$-dependent part of the supergravity BRST algebra decouples    by redefining
\begin{equation}   
\hat s=s- {\cal L}_\xi,\quad \hat d =d +\hat s  + i_\phi, 
\end{equation}
where the vector field  $\phi$  is a bilinear in the supersymmetry ghost $\chi$ 
\begin{equation} 
\label{phi}
\phi^\mu =-\frac{i}{2} \bar\chi \gamma^\mu \chi =s \xi^\mu -\xi^\nu\partial_\nu \xi^\mu,
\end{equation}
$i_V$ is the interior derivative on the manifold for a given vector $V$ and $\cal L$ is the Lie derivative, $\L_V = i_V d + d i_V$. One has the important property
\begin{equation} 
\hat d =\exp(-i_\xi)(d+s)\exp(+i_\xi) 
\end{equation}
which ensures that $(d+s)^2=0$ and ${\hat d}^2=0$ are equivalent, and $s^2=0  \Leftrightarrow   {\hat s}^2= {\cal L}_\phi$. The supergravity  BRST  transformations can be obtained by imposing  constraints on the curvatures~(\ref{curvatures}), in a way that merely generalizes  the Yang--Mills case. Using   ghost unification allows for a direct  check  of the off-shell closure by means of the  Bianchi identities. In the end, one finds the following action of the BRST operator $\hat s$ on the fields:
\begin{eqnarray}
\label{brs}
\sg e^a &=& -\Omega^{ab}e_b - i\bar{\chi}\gamma^a\lambda \CR 
\sg\lambda &=& -D\chi - \Omega^{ab}\gamma_{ab}\lambda - c\gamma^5\lambda \CR
\sg B_2 &=& -dB_{1}^1 - i\bar{\chi}\gamma^a\lambda e_a \\
\sg A &=& -dc - \tfrac{1}{2}i\bar{\chi}\gamma^5\gamma^aX_a \CR 
\sg \omega^{ab} &=& -(D\Omega)^{ab} - i\bar{\chi}\gamma^{[a}X^{b]} \nn
\end{eqnarray}
where the spinor $X_a$ is
\begin{equation}
\label{X}
X_a =\rho_{ab}e^b - \left(\tfrac{1}{2}G_{abc}\gamma^{bc} + \tfrac{1}{12}\epsilon_{abcd}G^{bcd}\gamma^5\right)\lambda
\end{equation}
The ghost transformation laws can be found in Appendix A. They are such that the closure relation $s^2=0  \Leftrightarrow   {\hat s}^2= {\cal L}_\phi$ is satisfied. The way the BRST symmetry transforms the supersymmetry ghost  will have non-trivial consequences  in the twisted formulation.
  
By using   the   twist formulas of  Majorana spinors  as in~\cite{lbN,Johansen,Witten1,popov,Hofman,bata}, one could analytically continue and  twist   by brute force these transformations in Euclidean space. 
 
We will rather try to obtain the twisted formulation in a more straightforward way, so as to unveil and better understand the mechanisms taking place in the twisted formalism. Therefore, we now proceed to our direct construction of the twisted superalgebra, keeping in mind that both untwisted and twisted formulations can be compared at any given stage.
  
As we will see, the whole information about supergravity is actually contained in the twisted  scalar nilpotent generator that  is hidden in the Poincar\'e supersymmetry algebra. To reach this result,   we need to separate both the Einstein and Rarita--Schwinger Lagrangians in parts depending only on the selfdual or the antiselfdual parts of the spin-connection.
  
\section{Selfdual decomposition of the supergravity action} 
 
Each of the Einstein and Rarita--Schwinger Lagrangians can be naturally split into two parts, one that only  depends on  the selfdual components of the spin-connection while the other one depends on the antiselfdual ones. These two parts are equal modulo suitable boundary terms.  In the case of the Einstein Lagrangian, this property was already used for other types of twisting~\cite{bata}.  
  
The Einstein Lagrangian can be written as~\footnote[1]{Our conventions for (anti)selfdual tensors are collected in Appendix~\ref{conventions}.}
\begin{equation}
L_E=\frac{1}{4}\epsilon_{abcd} e^a e^b R^{cd}= \frac{1}{2} e^a e^b \left(R^+_{ab} - R^-_{ab}\right)
\end{equation}
Since the \(so(4)\) Lie algebra splits into two parts, the selfdual components of the curvature $R^{\pm ab}= d\o^{\pm ab}  + \o^{\pm a}_{\:c}\o^{\pm cb}$ only depend  on the components of the   spin-connection $\o^{\pm ab}$ with the same selfduality. 

In supergravity, the torsion is often taken to be  $T_a=De_a +\frac{i}{2}  \bar \la\gamma_a\la$, but to establish the equality between the two parts of the Einstein Lagrangian, it is simpler to also use the purely bosonic torsion \(t_a \equiv De_a\) which satisfies the  Bianchi identity $Dt_a=R_{ab}e^b$. Indeed, contracting this identity with \(e^a\), one has:
\begin{equation}  
e^a Dt_a   =   e^a e^b (R^+_{ab} + R^-_{ab}) 
\end{equation}
while 
\begin{equation}
D(e^a t_a) = t^a t_a - e^a Dt_a
\end{equation}
One then gets: 
\begin{eqnarray}
L_E &=&  - e^a e^b R^-_{ab} + \frac 1 2 t^a t_a - \frac 1 2 d ( e^a t_a) \nonumber \\
&=& - e^a e^b R^-_{ab} -\frac{i}{2}   \bar \la \gamma ^a\la T_a +\frac{1}{2} {T^a} {T_a} - \frac{1}{2}d( e^aT_a  - \frac{i}{2}e^a  \bar \la \gamma_a\la ) \label{Einsteinasd} \\
&=& + e^a e^b R^+_{ab} + \frac{i}{2}   \bar \la \gamma ^a\la T_a - \frac{1}{2} {T^a} {T_a} + \frac{1}{2}d( e^aT_a  - \frac{i}{2}e^a  \bar \la \gamma_a\la ) \label{Einsteinsd}  
\end{eqnarray}
The second line is obtained by expressing \(t_a\) in terms of \(T_a\), remembering that \(\bar\la \gamma^a \la \bar \la \gamma_a \la = 0\) when \(\la\) is a Majorana spinor.

Since $T^a$ is constrained   to be zero or  a quantity independent of the   spin-connection, the expressions obtained for the Einstein action only  depend on the antiselfdual part $\o^{-ab}$ (in the case of Eq.~(\ref{Einsteinasd})) or the selfdual part  $\o^{+ab}$ (for Eq.~(\ref{Einsteinsd})) of the   spin-connection.

An  analogous  property  holds true  for the Rarita--Schwinger Lagrangian. One can derive it using the decomposition of the gravitino on its chiral components (which are not independent for a Majorana spinor). Defining     \(\la = \la^+ + \la^-\) with \(\la^\pm = \frac 12 (1 \pm i\gamma^5)\la \), one writes~\footnote[2]{See Appendix~\ref{conventions} for the details of our chirality conventions.}
\begin{equation}
L_{RS} = i\,\bar{\lambda}\gamma^{5}\gamma^a\rho e_a = \bar{\lambda}^{+} \gamma^a \rho^- e_a  -\bar{\lambda}^{-}\gamma^{a} \rho^{+} e_a
\end{equation}
using  $\bar\lambda^\pm\gamma^{a}\lambda^{\pm} = 0$~\footnote[3]{Care must be taken in Minkowski space where the conjugation changes chirality, so that for example \(\overline{\la^+} = \bar\la^-\).}. By adding a suitable total divergence, one gets    
\begin{equation}
\label{lagrangian_RS}
L_{RS} = 2 \bar\lambda^+\gamma^{a}\rho^{-}e_a - \bar\lambda^-\gamma^{a}\lambda^{+}T_a +\,d(\bar{\lambda}^{-}\gamma^{a}\lambda^{+}e_a )
\end{equation}
With anticommuting Majorana fermions, we have the identity \(\bar X^- \gamma_a Y^+ = - \bar Y^+ \gamma_a X^-\).  Since the chiral projections commute with the generators of Lorentz transformations on spinors, we simply have $\rho^-  = D(\la^-)$.  Chiral fermions give the minimal representations of the subalgebras associated to the selfdual and antiselfdual parts of the rotation generators, so that 
$\rho^-$ only depends on the antiselfdual part of the   spin-connection $\o^{-ab}$:
\begin{equation} 
\rho^- = \left(d + \tfrac{1}{2}\o^{-ab}\gamma_{ab} + iA\right)\la^{-}
\end{equation}

The Rarita--Schwinger action can therefore be written as  
\begin{equation}  
\label{RSsd} 
I_{RS}= i \, \int  \bar \la   \gamma^5\gamma^a  D^{( \omega ) }  \la  e_a = \int   { 2 \bar \la ^{  +}   \gamma^a D^{(   {  \omega}^  { -}) }  \la^{  {  -}} } e_a{  -  \bar \la^{-}  \gamma ^a   \la^{+}  } T_a
\end{equation}

We succeeded in expressing  $I_E+I_{RS}$   in a way that only depends on either the selfdual or the antiselfdual part of the    spin-connection, whenever the  constraint on the torsion is independent of the spin-connection. This condition is necessary for the closure of the supersymmetry algebra acting on the vielbein.   
   
\section {Twisted supergravity variables}

In order to be able to twist the theory, we must work in a Euclidean space with an almost complex structure, \textit{i.e.} a map on each tangent space $J(x)$ with $J^2=-1$, or more explicitly $J^{\mu}_{\rho}(x) J^{\rho}_{\nu}(x) = -\delta^{\mu}_{\nu}$. 

Introducing complex coordinates  $z_m, \bar z_{\bm}$, where $m=1,2$, one  can locally reduce  the complex structure to a diagonal one, $J_m^{\hphantom{m}n} = i\delta_m^{\hphantom{m}n}$, $J_\bm^{\hphantom{\bm}\bn} = -i\delta_\bm^{\;\;\;\bn}$. Making use of a compatible metric to lower one of the indices in \(J\), \(J\) becomes an  antisymmetric tensor with $J_{m\bn}$ as the only  non-vanishing components. 

The  tensor $J_{m\bn}$  can be used instead of the metric to lower and raise indices in the tangent space, according to $X^m = -iJ^{m\bn}X_\bn$ and $X^{\bm} = iJ^{\bm n}X_n$. In order to keep our formulas as uncluttered as possible, we will use a notation similar to Einstein's notation for contracting antiholomorphic and holomorphic $SU(2)$-indices by means of the complex structure constant tensor, as follows 
\begin{equation}
\label{formule}
X^{a}Y_{a} = g^{ab}X_{a}Y_{b} = -iJ^{m\bn}(X_mY_{\bn} + X_{\bn}Y_m) \equiv X_mY_\bm - X_\bm Y_m
\end{equation}
The antisymmetry of the tensor $J_{m\bn}$ implies that one must be careful about the ordering of indices. It  explains the minus sign appearing  in the last term of Eq.~(\ref{formule}).

Twisting must be done in Euclidean space where it is known that there are no Majorana spinors. We therefore forget the Majorana condition, the effect of which can be recovered afterwards from a careful consideration of the Wick rotation~\cite{nico}. We associate to the spinor $(\lambda^\alpha,\lambda_{\dot\alpha})$  the following four quantities with only holomorphic or antiholomorphic indices:
\begin{equation}
\label{deftwist} 
(\Psi_m, \Psi_{\bm\bn}, \Psi_0)
\end{equation}
The indices \(m\) and \(\bm\) take two different values and the object \(\Psi_{\bm\bn}\) is antisymmetric in its indices, so that it only has one non-zero component. 

The twisted components of a spinor~(\ref{deftwist}) are defined from the following linear mapping, which uses Pauli matrices elements~\cite{lbN, Johansen, Witten1, popov, Hofman, bata}:
\begin{eqnarray}
\label{twistn1}
\Psi_m &=& \lambda^\alpha(\sigma_m)_{\alpha\dot 1} \CR
\Psi_{\bm\bn} &=& \bar\lambda_{\dot\alpha}(\bar\sigma_{\bm\bn})_{\;\;\;\dot 2}^{\dot\alpha} \\
\Psi_0 &=& \bar\lambda_{\dot\alpha}\delta^{\dot\alpha}_{\dot 2} \nn  
\end{eqnarray}
In Appendix~\ref{twisted_gamma}, we give the expression of the twist of \(\Gamma \la\) as functions of the twisted components of \(\la\) for some elements \(\Gamma\) of the Clifford algebra.

This construction reduces the  tangent space $SO(4)$ symmetry into an $SU(2)\times U(1)\subset SO(4)$ symmetry.
With this change of variables,   $SO(4)$-invariant expressions can be related to their twisted counterparts, which generally  split into  a sum of independently $U(2)$-invariant terms.
For instance, the Rarita--Schwinger Lagrangian can be decomposed as follows
\begin{equation}
\bar{\lambda} \gamma^{5}\gamma^a\rho  e_a =(\Psi_0 \rho_m + \Psi_m \rho_0)  e_\bm - (2\Psi_{\bm\bn} \rho_n - \Psi_n \rho_{\bm\bn})  e_m
\end{equation}

The commuting  Majorana ghost of local supersymmetry $\chi$ is  twisted as follows
\begin{equation}
\chi \sim (\chi_m,\chi_{\bm\bn},\chi_0)
\end{equation}
and the vector field in Eq.~(\ref{phi}), $\phi^\mu  = -\tfrac{i}{2} \bar\chi   \gamma^\mu \chi= s\xi^\mu-\xi^\nu \partial_\nu \xi^\mu$  is now given by 
\begin{equation}
\label{vect_phi}
\phi_m = -\chi_m \chi_0 , \quad  \phi_\bm = -\chi_{\bm\bn}\chi_n
\end{equation}
When the parameter of vector supersymmetry vanishes, $\chi_m=0$, then  the vector field $\phi$ vanishes~\footnote[4]{This condition means that $\chi$ is a pure spinor, and it is not surprising  that it entails great simplifications in the formalism, as in~\cite{berko}.}.

A consistent interpretation of  the  twisted supersymmetry   only involves   fermionic  global charges. Thus, in what follows,     $\chi_m,\chi_{\bm\bn},\chi_0$ will be treated  as constant ghosts.   We will build a  set of corresponding generators $\delta_\bm,\delta_{mn},\delta$ that satisfy anticommutation relation that close independently of the equations of motion (off-shell closure), but possibly modulo bosonic gauge transformations.  We will consider  the operation
\begin{equation}
Q=    \chi_m  \delta_\bm+     \chi_{\bm\bn}  \delta_{mn}   +  \chi_0 \delta
\end{equation}
For a vanishing  gravitino  field,  $Q$ is nilpotent,  off-shell and  modulo   bosonic gauge transformations.  It turns out that  the  global  $Q$-invariance   is a sufficiently strong condition  to  determine the supergravity action. In fact, it gives a Ward identity that is sufficient to control the quantum perturbative behavior of the theory generated by the $Q$-invariant action, once all its gauge invariances are gauge-fixed in a BRST invariant way. When  the gravitino  field is not zero, the closure algebra is more involved. We will see that it involves  supersymmetry transformations with gravitino field dependent structure coefficients. 

In this construction, the supergravity action  is   however fully  determined by the global supersymmetry operation $Q$. Local supersymmetry is  warranted due to the systematic construction of the charges  $\delta_\bm$, $\delta_{mn}$, $\delta$   in a way that is compatible with the Bianchi identities of all field curvatures.

The four generators $(\s,\s_\bm,\s _{mn})$   must act on all the twisted fields of the multiplet~(\ref{multiplet}), with the following $g$-grading assignments. 
\begin{eqnarray}
&
\begin{array}{|c|c||c|c|}\hline
\: \mbox{Field} \: & \: \mbox{Grading} \: & \: \mbox{Field} \: & \: \mbox{Grading} \:  \\ \hline
e_m & 0  & A & 0  \\
e_\bm & 0  & B_2 & 0  \\ \hline
\Psi_m & 1  & \o_{mn} & 0  \\
\Psi_0 & -1  & \o_{\bm\bn} & 0  \\
\Psi_{\bm\bn} & -1  & \o_{m\bn} & 0  \\ \hline
\end{array}
&\CR \nn
&
\begin{array}{|c|c|}\hline
\:\mbox{Generators} \: & \: \mbox{Grading} \: \\ \hline
\s & 1    \\
\s_\bm & -1    \\
\s_{mn} & 1    \\ \hline
\end{array}
&\CR \nn
\end{eqnarray}
 The commutation properties of the various fields are always obtained by computing the   sum of the form degree and the grading  $g$ of fields (for instance $e_m$ is an anticommuting object  since the form degree is  one and $g=0$, $\Psi _m$ is a commuting object  since the form degree is  one and $g=1$, etc.). After having obtained a  classical action that is invariant under the twisted nilpotent global supersymmetry $Q$, one must in principle check that it remains  invariant under local supersymmetry by giving a coordinate dependence to    $(\chi_0,  \chi_m, \chi_{\bm\bn})$. This is in fact automatically realized, since all derivatives will appear as  super-covariantized ones.

If we now generalize   $(\chi_0,  \chi_m, \chi_{\bm\bn})$ into local commuting  (twisted) Faddeev Popov ghosts, one gets the     operator 
\begin{equation}
\hat s = \chi_0(x)\delta + \chi_m(x)\delta_\bm + \chi_{\bm\bn}(x)\delta_{mn} 
\end{equation}
Its action on the classical fields is the same as that  of the  standard  BRST transformations     in twisted form.
 
  In   the flat space $N=1$ super-Yang--Mills theory~\cite{lbN}, the three nilpotent symmetry generators $\delta$ and $\delta_\bp$  satisfy 
the   off-shell closure anticommutation relations  $\delta^2 = 0$,  $\{\delta_{\bp},\delta_{\bq}\} = 0$, $\{\delta,\delta_\bp\} = \partial _\bp$.

The situation is more complicated in supergravity.   In this case, one has  indeed 
the property $  \hat s ^2= {\cal L}_\phi$,  where the vector field $\phi$ has been  defined in~(\ref{vect_phi}).   One has also    the transformation law $\hat s \chi \sim i_\phi \Psi$ (see Appendix A),   which  remains true even when the supersymmetry ghosts are assumed to be constant. 
This implies the following supergravity generalization
\bea
\label{anticomm}
\delta^2 &=& 0, \quad \quad \quad \quad \quad \quad \quad \quad \{\delta_{\bp},\delta_{\bq}\} = 0, \CR
\{\delta,\delta_\bp\} &=& {\cal L} _\bp - \displaystyle\sum\limits_{a= 0,m,\bm\bn}\Psi_{\bp,a}\delta_\ba
\eea

These anticommutation relations hold modulo bosonic gauge transformations.  The   derivative   ${\cal L} _\bp$ is     the Lie derivative along the   vector field dual to the vielbein component $e_\bp $.

 In fact, the supersymmetry generators that occur  in the expression of $\{\delta,\delta_\bp\}$ occur proportionaly to the   components $\Psi_{\bp,a}$ of the gravitino field $\Psi_a \equiv \Psi_{m,a}e_\bm - \Psi_{\bm,a}e_m$.

One  thus recognizes the expected  feature of supergravity:  the anticommutator $\{\delta,\delta_\bp\}$ closes on supersymmetry generators with field-dependent coefficients, proportionally to   gravitino-field components.

Therefore, it is expected that the  anticommutator  $\{\delta,\delta_{\bq}\}$  involves   the fourth symmetry generator $\delta_{pq}$, whose existence can  be  checked afterwards in the twisted  method.

Once $\delta$ and $\delta_\bp$ are determined, the $\delta$ and $\delta_\bp$ invariant action turns out to be automatically invariant under a $\delta_{pq}$ symmetry. In the four-dimensional supergravity, the relations between $\delta_{pq}$ and the other generators $\delta$ and $\delta_\bp$ are satisfied off-shell. 

 The fermionic  scalar operator $\delta$  can be extended as a   globally well-defined  object (provided there is a complex structure). We   will mainly focus on the question of its direct construction. In  fact, $\delta_\bp$  and $\delta_{pq}$ can only be given a geometrical interpretation on a coordinate patch. 

\section{The supergravity  curvatures in the $U(2)\subset  SO(4)$ invariant formalism}
  
In section~3, we have shown that both the  Einstein and Rarita--Schwinger actions only depend on the selfdual or antiselfdual components of the spin-connection. 
In $SU(2)\subset  SO(4)$ notations,  the  selfduality condition  of an antisymmetric Lorentz tensor, $F_{ab}= \tfrac{1}{2}\epsilon_{abcd}F^{cd} $ reads
\begin{equation}
F_{\bm\bn}= F_{mn}=0     \ \ \ \  F_{m\bm} \equiv iJ^{m \bn}F _{m\bn}=0
\end{equation}
while the   antiselfduality condition $F_{ab}= -\tfrac{1}{2}\epsilon_{abcd}F^{cd} $ reads   
\begin{equation} 
F_{m\bn} - iJ_{m\bn} F_{p\bp} =0 
\end{equation}
Thus, the    spin-connection $\omega^{ab}=\omega^{+ab} +\omega^{-ab}   \equiv  (\o_{mn},\o_{\bm\bn},\o_{m\bn} )  $  splits in selfdual and antiselfdual  parts, respectively
\begin{eqnarray}
\o^{+ab}  &\sim &   (0,0,\o_{m\bn} - iJ_{m\bn}  \o ) \CR
\o^{-ab}    &\sim&   (\o_{mn},\o_{\bm\bn},iJ_{m\bn}\o )  
\end{eqnarray}
where $ \o\equiv iJ_{\bm n} \o_{m\bn}$.

The $SO(4)$ Lie algebra is the product of two \(SU(2)\) corresponding to the selfdual and antiselfdual generators.  Therefore, the antiselfdual part of the curvature 2-form $R^- \sim (R_{mn}, R_{\bm\bn}, R)$ and its  Bianchi identities only depend on the antiselfdual part of the connection \(\o^{-ab}\):
\begin{eqnarray}
\label{curv_R}
&&R = d\o + 2\o_{mn}\o_{\bm\bn}  \CR
&&R_{mn} = d\o_{mn} - \o\o_{mn}  \CR
&&R_{\bm\bn} = d\o_{\bm\bn} + \o\o_{\bm\bn} \\
&&dR = 2R_{mn}\o_{\bm\bn} - 2\o_{mn}R_{\bm\bn} \CR
&&dR_{mn} = R_{mn}\o - R\o_{mn} \CR
&&dR_{\bm\bn} = R\o_{\bm\bn} - R_{\bm\bn}\o \nn
\end{eqnarray}
The $SO(4)$ symmetry only  acts as this $SU(2)$ on $ \Psi_0$ and $\Psi_{\bm\bn}$, due to chirality properties.   One can thus   define the $SU(2)$ covariant   curvatures  for  $ \Psi_0$ and $\Psi_{\bm\bn}$ 
\begin{eqnarray}
\rho_0 &=& d\Psi_0 - \Bigl(\frac{1}{2}\o - A\Bigr)\Psi_0 + \o_{mn}\Psi_{\bm\bn} \CR
\rho_{\bm\bn} &=& d\Psi_{\bm\bn} + \Bigl(\frac{1}{2}\o + A\Bigr)\Psi_{\bm\bn} - \o_{\bm\bn}\Psi_0
\end{eqnarray}
Their Bianchi identities are
\begin{eqnarray}
D\rho_0 &=& \Bigl(-\frac{1}{2}R + F\Bigr)\Psi_0 + R_{mn}\Psi_{\bm\bn} \CR
D\rho_{\bm\bn} &=& \Bigl(\frac{1}{2}R + F\Bigr)\Psi_{\bm\bn} - R_{\bm\bn}\Psi_0
\end{eqnarray}
The curvature  $\rho _m $ of $\Psi_m$ only  involves the selfdual part of the spin-connection.  We can skip its definition, since   it is not needed in  the supergravity action.

The  torsion involves both selfdual and antiselfdual  components of the  spin-connection  
\begin{eqnarray}
\label{Bianchi_T}
T_m&=&de_m +\o_{mn}e_\bn - \o_{m\bn}e_n + \Psi_m\Psi_0       \CR
T_\bm&=&de_\bm +  \o_{\bm n}e_\bn - \o_{\bm\bn}e_n + \Psi_{\bm\bn}\Psi_n \\
DT_m &=& R_{mn}e_\bn - R_{m\bn}e_n + \rho_m\Psi_0 - \Psi_m\rho_0 \CR
DT_\bm &=& R_{\bm n}e_\bn - R_{\bm\bn}e_n  + \rho_{\bm\bn}\Psi_n - \Psi_{\bm\bn}\rho_n
\end{eqnarray}

We  now use the  $SU(2)$  notation to decompose  the Einstein and Rarita--Schwinger Lagrangians as a sum of terms that are separately $SU(2)$ invariant,  using  the  expressions~(\ref{Einsteinasd}) and~(\ref{RSsd}) 
\begin{eqnarray}
\label{lagrangian_E_twisted}
I_E &=& \int  -\Bigl(Re_me_\bm + R_{mn}e_\bm e_\bn + R_{\bm\bn}e_me _n\Bigr) - \Bigl(\Psi_m\Psi_0T_\bm - \Psi_{\bm\bn}\Psi_nT_m\Bigr) + T_m T_\bm \\
\label{lagrangian_RS_twisted}
I_{RS} &=& \int - \Bigl(2 \rho_{\bm\bn}\Psi_ne_m - 2\rho_0\Psi_me_\bm\Bigr) + \Bigl(\Psi_m\Psi_0T_\bm  - \Psi_{\bm\bn}\Psi_nT_m\Bigr)
\end{eqnarray}
Eqs.~(\ref{lagrangian_E_twisted}) and (\ref{lagrangian_RS_twisted}) are interesting. However, at first sight,  they are not yet very suggestive about the existence of a twisted scalar supersymmetry.

In fact, to build the scalar supersymmetry, we  depart  from the method used in~\cite{baubel}.  The so-called 1.5 order formalism,  once adapted to the twisted fields  of supergravity, will  neatly separate  the various  terms of  the invariant actions~(\ref{lagrangian_E_twisted}) and (\ref{lagrangian_RS_twisted}). 

\section{1.5 order formalism with $SU(2)$ covariant curvatures}

The   justification of the  1.5 order  formalism for supergravity is detailed in \cite{PVN}. One first builds a supersymmetry that acts on all fields but the spin-connection $\o$. The later is  taken not to transform  under supersymmetry in a first step. 

The second order formalism transformation law of $\o$ is the one compatible with all  Bianchi identities  of the theory, including that of the Riemann curvature.  
 
In the  1.5 order formalism, it  is  particularly simple  to obtain the twisted scalar supersymmetry on all fields but the spin-connection, by imposing consistent constraints on the ghost-dependent curvatures.

The ghost-dependent  curvatures are obtained   by  the substitutions
\begin{equation}
\label{subs}
d \to \hat d = d +\chi_0  \sos + i_\phi , \quad  \Psi \to \hat\Psi = \Psi + \chi
\end{equation}
We are only concerned with the scalar supersymmetry for the moment.  Thus, we only retain a constant $\chi_0$    as the  only non-vanishing component in $\chi$. Since  $\chi_m=0$, one has   $\phi_m = \phi_\bm = 0$ and    $\hat d = d + \chi_0\sos$. The property    ${\hat d}^2=0$   implies  $\sos^2 = 0$ on all fields. The 1.5 order formalism  constraints  that are compatible with the Bianchi identities are 
\begin{eqnarray}\label{brst1,5}
\hat R &=& R \quad \hat R_{mn} = R_{mn} \quad \hat R_{\bm\bn} = R_{\bm\bn} \quad \hat F = F \CR
\hat\rho _0 &=& \rho_0 \quad \hat\rho_{\bm\bn} = \rho_{\bm\bn} \quad \hat\rho_{m} = \rho_{m} \\
\hat G_3 &=& G_3 \quad \hat T_m = T_m \quad \hat T_\bm = T_\bm \nn
\end{eqnarray}
where $G_3$ is the field strength of the 2-form $B_2$, defined in twisted form as\begin{equation}\hat G_3 = \hat dB_2 + \hat\Psi_m \hat \Psi_0e_\bm - \hat\Psi_{\bm\bn} \hat\Psi_n e_m
\end{equation}
We now use Eq.~(\ref{subs})  and pick up the term  with    ghost number one in Eq.~(\ref{brst1,5}). This gives  the $\sos$-transformation laws for all fields:
\begin{equation}
\label{on-shell}
\begin{array}{|c||c|} \hline
&\sos \\ \hline\hline
e_m & -\Psi_m \\
e_\bm & 0 \\ \hline
\Psi_m & 0 \\
\Psi_0 & \: \: \tfrac{1}{2}\o-A \: \: \\
\Psi_{\bm\bn} & \o_{\bm\bn} \\ \hline 
\o_{mn} & 0 \\
\: \: \o_{\bm\bn} \: \: & 0 \\
\o & 0 \\ \hline
A & 0 \\ \hline
B_2 & -\Psi_m e_\bm \\ \hline
\end{array}
\end{equation}
The  curvatures transform as
\begin{eqnarray}
\sos R &=& 0 \quad \quad \sos R_{mn} = 0 \quad \quad \sos R_{\bm\bn} = 0 \quad \quad \sos F = 0 \CR
\sos \rho_0 &=& -\tfrac{1}{2}R + F \quad \quad \sos \rho_{\bm\bn} = -R_{\bm\bn} \quad \quad \sos \rho_m = 0
\end{eqnarray}
We can therefore build three $\sos$-invariant Lagrangians that  respectively  contain the three independent $SU(2)$-invariant pieces $Re_me_\bm$,    $   R_{mn}e_\bm e_\bn $ and   $R_{\bm\bn}e_me _n $ of the Einstein Lagrangian: 
\begin{eqnarray}
&&R_{mn}e_\bm e_\bn   \CR
&&R_{\bm\bn}e_me_n + 2\rho_{\bm\bn}\Psi_ne_m \\
&&Re_me_\bm - 2\rho_0\Psi_me_\bm - 2FB_2 \nn   
\end{eqnarray}
The action
\begin{equation}
\label{action_on-shell}
I = \int\alpha R_{mn}e_\bm e_\bn + \beta(R_{\bm\bn}e_me_n + 2\rho_{\bm\bn}\Psi_ne_m) + \gamma(Re_me_\bm - 2\rho_0\Psi_me_\bm - 2FB_2)
\end{equation}
is  thus invariant under the transformations~(\ref{on-shell}), for all possible values of the coefficients $\alpha$, $\beta$ and $\gamma$. Lorentz symmetry  is obtained when $\alpha=\beta=\gamma$.

Alternatively, in a method that is closer to the one used in~\cite{baubel}, one  can  directly check the invariance of the action~(\ref{action_on-shell}) by computing the following quantities, using the Bianchi identities for the curvatures:
\begin{eqnarray}
\label{derivatives}
\hat{D}(\hat{R}_{mn}\hat{e}_\bm\hat{e}_\bn) &=& 2\hat{R}_{mn}(\hat{T}_\bm - \hat{\Psi}_{\bm\bp}\hat{\Psi}_p)\hat{e}_\bn \CR
\hat{D}(\hat{R}_{\bm\bn}\hat{e}_m\hat{e}_n) &=& 2\hat{R}_{\bm\bn}(\hat{T}_m - \hat{\Psi}_m\hat{\Psi}_0)\hat{e}_n \CR
\hat{D}(\hat{R}\hat{e}_m\hat{e}_\bm) &=& \hat{R}(\hat{T}_m - \hat{\Psi}_m\hat{\Psi}_0)e_\bm - \hat{R}\hat{e}_m(\hat{T}_\bm - \hat{\Psi}_{\bm\bp}\hat{\Psi}_p) \\
\hat{D}(\hat\rho_{\bm\bn}\hat\Psi_n\hat{e}_m) &=& \Bigl(\bigl(\frac{1}{2}\hat{R} + \hat{F}\bigr)\hat{\Psi}_{\bm\bn} - \hat{R}_{\bm\bn}\hat{\Psi}_0\Bigr)\hat\Psi_n\hat{e}_m + \hat{\rho}_{\bm\bn}\hat\rho_n\hat{e}_m - \hat{\rho}_{\bm\bn}\hat{\Psi}_n(\hat{T}_m - \hat{\Psi}_m\hat{\Psi}_0) \CR
\hat{D}(\hat\rho_0\hat\Psi_m\hat{e}_\bm) &=& \Bigl(\bigl(-\frac{1}{2}\hat{R} + \hat{F}\bigr)\hat{\Psi}_0 + \hat{R}_{pq}\hat{\Psi}_{\bp\bq}\Bigr)\hat\Psi_m\hat{e}_\bm + \hat{\rho}_0\hat{\rho}_m\hat{e}_\bm - \hat{\rho}_0\hat{\Psi}_m(\hat{T}_\bm - \hat{\Psi}_{\bm\bp}\hat{\Psi}_p) \CR
\hat{D}(\hat{F}\hat{B}_2) &=& \hat{F}(\hat{G}_3 - \hat{\Psi}_m\hat{\Psi}_0\hat{e}_\bm + \hat{\Psi}_{\bm\bn}\hat{\Psi}_n\hat{e}_m) \nn
\end{eqnarray}
Taking the part with ghost number 1 of these equations and retaining only $\chi_0 \neq 0$, one obtains the $\sos$ transformations of the various terms in the action:
\begin{eqnarray}
\sos(R_{mn}e_\bm e_\bn) &=& 0 \CR
\sos(R_{\bm\bn}e_me_n) &=& -2R_{\bm\bn}\Psi_m e_n \CR
\sos(Re_me_\bm) &=& -R\Psi_me_\bm \\
\sos(\rho_{\bm\bn}\Psi_ne_m) &=& -R_{\bm\bn}\Psi_ne_m \CR
\sos(\rho_0\Psi_me_\bm) &=& \bigl(-\tfrac{1}{2}R + F\bigr)\Psi_me_\bm \CR
\sos(FB_2) &=& -F\Psi_m e_\bm \nn
\end{eqnarray}
which ensure that $\sos(I) = 0$. 

The formulas~(\ref{derivatives}) are actually quite useful to directly compute   the action of the vector supersymmetry $\s^{1.5}_\bp$, by generalizing to the case where  $\chi_p \neq 0$. Using a ghost  expansion as  for the scalar symmetry, one gets  
\begin{eqnarray}
\s^{1.5}_\bp(R_{mn}e_\bm e_\bn) &=& 2R_{mn}\Psi_{\bm\bp}e_\bn \CR
\s^{1.5}_\bp(R_{\bm\bn}e_me_n) &=& 2R_{\bp\bn}\Psi_0 e_n \CR
\s^{1.5}_\bp(Re_me_\bm) &=& -R\Psi_0e_\bp + Re_m\Psi_{\bm\bp} \\
\s^{1.5}_\bp((\rho_{\bm\bn}\Psi_ne_m) &=&\Bigl(- \bigl(\frac{1}{2}R + F\bigr)\Psi_{\bm\bp} + R_{\bm\bp}\Psi_0\Bigr)e_m \CR
\s^{1.5}_\bp(\rho_0\Psi_me_\bm) &=& \Bigl(\bigl(-\frac{1}{2}R + F\bigr)\Psi_0 + R_{mn}\Psi_{\bm\bn}\Bigr)e_\bp \CR
\s^{1.5}_\bp(FB_2) &=& -F(\Psi_0 e_\bp + \Psi_{\bm\bp}e_m) \nn
\end{eqnarray}
One finds that  $\s^{1.5}_\bp$ is  another symmetry of the complete action,  provided that $\alpha=\beta=\gamma$, in which   case   the $SU(2)$ symmetry is enlarged to   $SO(4)$.

However, one must be careful in the interpretation of   this vector symmetry, since  it cannot be  obtained by  twisting the supersymmetry generators $\left(Q^\alpha, Q_{\dot{\alpha}}\right)$. Indeed,   $\sos$ and $ \s^{1.5}_\bp$ do not have the right anticommutation relations,    since $\{\sos, \s^{1.5}_\bp\}\Psi = 0$,  in contradiction with the twisted supersymmetry algebra~(\ref{anticomm}). In fact the 1.5 order formalism, which is useful to determine the invariant action, does not properly define the supersymmetry generators.  One must determine the $\o$ transformations consistent with the constraints, which appear as equations of motion in the 1.5 order formalism.

With the invariant action~(\ref{action_on-shell}), the equations of motion of the  antiselfdual spin-connection give  $12 = 3 \times 4$ equations that  can be solved algebraically to determine  the 12 components of the three 1-forms $\omega_{mn}$, $\omega_{\bm\bn}$ and $\o$, as functions of $e$ and $\Psi$. The precise values then depend on the parameters $\alpha,\beta$ and $\gamma$. 
 
One can then compute the  $\sos$ transformations of these functions through the chain rule  to obtain the transformations of  $\omega_{mn}$, $\omega_{\bm\bn}$, $\o$. Since $\sos$ is nilpotent on $e$ and $\Psi$,  this procedure  gives a nilpotent transformation in the second order formalism, where $\omega_{mn},\omega_{\bm\bn}$ and $\o$ are not independent fields.
 
The case of interest is  for the rotationally invariant action~(\ref{action_on-shell}), which has  $\alpha=\beta=\gamma $. In this case, the spin-connection equations of motion give
\begin{eqnarray}
\frac{\s }{\s\o} I(e,\Psi,B_2,\o) &=& e^{\vphantom {(\o^-)}}_{m}T^{(\o^-)}_{\bm} = 0 \CR
\frac{\s}{\s\o_{mn}}  I(e,\Psi,B_2,\o) &=& e^{\vphantom {()\o^-}}_{[\bm}T^{(\o^-)}_{\bn]} = 0 \\
\frac{\s }{\s\o_{\bm\bn}} I(e,\Psi,B_2,\o) &=& e^{\vphantom {()\o^-}}_{[m}T^{(\o^-)}_{n]} = 0 \nn
\end{eqnarray}
Here $T^{(\o^-)}$ is a function only of $\o^-$,  
\begin{eqnarray}
T^{(\o^-)}_m &=& de_m + \o_{mn}e_\bn + \Psi_m\Psi_0 \CR
T^{(\o^-)}_\bm &=& de_\bm - \o_{\bm\bn}e_n + \Psi_{\bm\bn}\Psi_n \nn
\end{eqnarray}
These 12 equations fix  the 12 components of the antiselfdual part of the spin-connection, $\o = \o(e,\Psi)$, $\o_{mn} = \o_{mn}(e,\Psi)$ and  $\o_{\bm\bn} = \o_{\bm\bn}(e,\Psi)$, as functions of the vielbein and the twisted gravitino. These components are the antiselfdual parts of the  complete spin-connection which satisfy the constraint $T_m=T_\bm=0 $,

As a consequence of the chain rule, $ \o(e,\Psi)$, $ \o_{mn}(e,\Psi) $ and $\o_{\bm\bn}(e,\Psi)$ transform under supersymmetry, and the 1.5 formalism guarantees that 
\begin{equation}
\label{action_off-shell_T_null}
I = -\int R_{mn}e_\bm e_\bn + (R_{\bm\bn}e_me_n + 2\rho_{\bm\bn}\Psi_ne_m) + (Re_me_\bm - 2\rho_0\Psi_me_\bm - 2FB_2)
\end{equation}
is still supersymmetric. 

To avoid the heavy calculations from the chain rule, one can use the formalism used in~\cite{baubel} and determine modified horizontality conditions for the field strengths $\hat R$ and $\hat F$ at  ghost numbers 1 and 2, such that the Bianchi identities are satisfied and the constraints are invariant. The invariance of the constraints is equivalent to the satisfaction of the chain rule. One defines
\begin{eqnarray}
\label{Rhoriz}
\hat{R} &=& R + R^{(1)} + R^{(2)} \\
\label{Fhoriz}
\hat{F} &=& F + F^{(1)} + F^{(2)} 
\end{eqnarray}
while we keep
\begin{eqnarray}
\hat{T} &=& T \CR
\hat{\rho} &=& \rho \\
\hat{G_3} &=& G_3 \nn
\end{eqnarray}
The ghost number two part of the Bianchi identity on the torsion $\hat T$ ensures that, when $\chi_m = 0$, $R^{(2)}= F^{(2)}=0$. The condition $\hat{G_3} =G_3 $ implies $\s B_2  = -\Psi_m e_\bm$ and
\begin{eqnarray}
\hat{\rho}_0 &=& (d+s)\hat{\Psi}_0 - \bigl(\demi\hat{\o} - \hat{A}\bigr)\hat{\Psi}_0 + \hat{\o}_{mn}\hat{\Psi}_{\bm\bn} = \rho_0 \CR
\hat{\rho}_{\bm\bn} &=& (d+s)\hat{\Psi}_{\bm\bn} + \bigl(\demi\hat{\o} + \hat{A}\bigr)\hat{\Psi}_{\bm\bn} - \hat{\o}_{\bm\bn}\hat{\Psi}_0 = \rho_{\bm\bn}
\end{eqnarray}
together with their respective Bianchi identities imply  
\begin{eqnarray} 
R^{(1)} &=& 2F^{(1)} \CR
R^{(1)}_{\bm\bn} &=& 0
\end{eqnarray}
Finally, the part with ghost number 1 of the Bianchi identity on $\hat T$~(\ref{Bianchi_T}) implies
\begin{eqnarray}
R^{(1)}_{mn} &=& -\tfrac{1}{2}\left(\rho_{p[n,m]}e_\bp + \rho_{\bp [n,m]}e_p\right) \CR
R^{(1)} &=& \left(\rho_{\bp\bm,m}e_p + \rho_{p\bm,m}e_\bp\right)
\end{eqnarray}
These values of $R^1$ and $F^1$ determine the transformation laws of $\o$ and $A$, so that the second order scalar supersymmetry transformations that  leave invariant the action~(\ref{action_off-shell_T_null}) are 
\begin{equation}
\begin{array}{|c||c|} \hline
&\delta \: (\textnormal{with} \: \delta^2=0) \\ \hline\hline
e_m & -\Psi_m \\
e_\bm & 0 \\ \hline
\Psi_m & 0 \\
\Psi_0 & \tfrac{1}{2}\o-A \\
\Psi_{\bm\bn} & \o_{\bm\bn} \\ \hline 
\o_{mn} & \: -\tfrac{1}{2}\left(\rho_{p[n,m]}e_\bp + \rho_{\bp [n,m]}e_p\right) \: \\
\o_{\bm\bn} & 0 \\
\o & \rho_{\bp\bm,m}e_p + \rho_{p\bm,m}e_\bp \\ \hline
A & \tfrac{1}{2}\left(\rho_{\bp\bm,m}e_p + \rho_{p\bm,m}e_\bp\right) \\ \hline
B_2 & -\Psi_m e_\bm \\ \hline
\end{array}
\end{equation}

We used a notation where $\rho_{mn}$, $\rho_{\bm\bn}$ and $\rho_{m\bn}$ are the components of the two-form $\rho$ on the vielbein basis, \textit{i.e.}, $\rho = \tfrac{1}{2}\left(\rho_{mn}e_\bm e_\bn + \rho_{\bm\bn}e_me_n + \rho_{\bm n}e_me_\bn\right)$.  The indices on the right  of the comma refer to the  twisted spinor indices $0$, $m$  or $\bm\bn$.

\section{Vector supersymmetry and non-vanishing torsion}

 {There is  no  vector supersymmetry  $\s_\bp$  for the action~(\ref{action_off-shell_T_null}) that can satisfy   the off-shell closure relation $\{\s, \s_\bp\} = {\cal L}  _\bp - \Psi_{\bp,a}\delta_\ba$}. Indeed, suppose that  such a symmetry exists. The off-shell closure means    ${\hat d}^2=   (d + \chi_0\s + \chi_p \d_\bp +  i_\phi)^2 =0$, with  $\phi_m = -\chi_m\chi_0 \neq 0$. Thus, the Bianchi identity,
\begin{equation}
\hat d G_3 = - \hat{\Psi}_m\hat{\rho}_0\hat{e}_\bm + \hat{\rho}_m\hat{\Psi}_0\hat{e}_\bm + \hat{\Psi}_m\hat{\Psi}_0\hat{T}_\bm - \hat{\rho}_{\bm\bn}\hat{\Psi}_n\hat{e}_m + \hat{\Psi}_{\bm\bn}\hat{\rho}_n\hat{e}_m - \hat{\Psi}_{\bm\bn}\hat{\Psi}_n\hat{T}_m
\end{equation}
has a non-trivial ghost number 2 part,  which is
\begin{equation}
\label{rel}
i_{\phi}  G_3 = \chi_m\chi_0 T_\bm.
\end{equation}
Therefore, the torsion cannot be taken identically equal to zero, which implies that the Lagrangian found in the previous section must be modified by terms that have an off-shell relevance. To remain in the context of a Lorentz invariant action, we use the following constraints on the torsion, which generalize Eq.~(\ref{rel}):
\begin{eqnarray}
\label{torsion_non_zero}
T_m &=& de_m +\o_{mn}e_\bn - \o_{m\bn}e_n + \Psi_m\Psi_0 = \tfrac{1}{2}\left(G_{m\bp\bq}e_pe_q - G_{mp\bq}e_\bp e_q\right) \CR
T_\bm &=& de_\bm + \o_{\bm n}e_\bn - \o_{\bm\bn}e_n + \Psi_{\bm\bn}\Psi_n = \tfrac{1}{2}\left(G_{\bm pq}e_\bp e_\bq - G_{\bm\bp q}e_p e_\bq\right)
\end{eqnarray}
The value of the spin-connection is therefore changed and the distortion on the horizontality condition~(\ref{Rhoriz}) becomes:
\begin{eqnarray}
R^{(1)}_{mn} &=& -\tfrac{1}{2}\left(\rho_{p[n,m]}e_\bp + \rho_{\bp [n,m]}e_p + G_{mn\bp}\Psi_p\right) \CR
R^{(1)} &=& \rho_{\bp\bm,m}e_p + \rho_{p\bm,m}e_\bp + G_{m\bp\bm}\Psi_p
\end{eqnarray}
The scalar supersymmetry transformations are now:
\begin{equation}
\label{table_off-shell}
\begin{array}{|c||c|}\hline
& \delta \: (\textnormal{with} \: \delta^2=0) \\ \hline\hline
e_m & -\Psi_m \\
e_\bm & 0 \\ \hline
\Psi_m & 0 \\
\Psi_0 & \tfrac{1}{2}\o-A \\
\Psi_{\bm\bn} & \o_{\bm\bn} \\ \hline 
\: \o_{mn} \: & \: -\tfrac{1}{2}\left(\rho_{p[n,m]}e_\bp + \rho_{\bp [n,m]}e_p + G_{mn\bp}\Psi_p\right) \: \\
\o_{\bm\bn} & 0 \\
\o & \: \rho_{\bp\bm,m}e_p + \rho_{p\bm,m}e_\bp + G_{m\bp\bm}\Psi_p \: \\ \hline
A & \: \tfrac{1}{2}\left(\rho_{\bp\bm,m}e_p + \rho_{p\bm,m}e_\bp + G_{m\bp\bm}\Psi_p\right) \: \\ \hline
B_2 & -\Psi_m e_\bm \\ \hline
\end{array}
\end{equation}
With $ T\neq 0$, the variation of the action found in the previous section involves new terms proportional to \(T \delta \o \), with must be compensated by the variation of new terms quadratic in \(G\).

One has
\begin{eqnarray}
&&\s G_{mp\bq} = \rho_{\bq p,m} - G_{mp\br}\Psi_{\bq,r} - 2G_{m\br\bq}\Psi_{p,r} \CR
&&\s G_{m\bp\bq} = \rho_{\bp\bq,m} - 2G_{m\br\bq}\Psi_{\bp,r} \\
&&\s e = -\tfrac{1}{2}\epsilon_{\bp\bn rs}\Psi_pe_ne_\br e_\bs \CR
&&\s({}^*G_3G_3) = -eG_{\bm pq}\Bigl(G_{r\bp\bq}\Psi_{\br,m} + 2G_{m\br\bq}\Psi_{\bp,r}\Bigr) 
\end{eqnarray}
Here ${}^*G_3$ denotes the Hodge dual of $G_3$ and \(e\) is the volume form built from \((e_m,e_\bm)\).

From the relation between the torsion and the 3-form $G_3$, Eq.~(\ref{rel}) one has:
\begin{equation}
T_mT_\bm + {}^*G_3G_3 = - \frac{1}{4}\left(G_{m\bp\bq}G_{\bm\br s}e_pe_qe_re_\bs + G_{mp\bq}G_{\bm rs}e_\bp e_qe_\br e_\bs\right)
\end{equation}
We thus add the term $T_mT_\bm + {}^*G_3G_3$ to the action~(\ref{action_off-shell_T_null}), which  cancels the effect of the variations of the spin-connection given in~(\ref{table_off-shell}) under the $\s$ symmetry. The resulting invariant action  is
\begin{equation}
I_{\textnormal{tot}} = -\int R_{mn}e_\bm e_\bn + (R_{\bm\bn}e_me_n + 2\rho_{\bm\bn}\Psi_ne_m) + (Re_me_\bm - 2\rho_0\Psi_me_\bm - 2FB_2) - T_mT_\bm - {}^*G_3G_3 
\end{equation}
Using Eqs,\ (\ref{lagrangian_E_twisted}) and~(\ref{lagrangian_RS_twisted}), this action can be written as
\begin{equation}
I_{\textnormal{tot}} = \int L_E + L_{RS} + 2FB_2 + {}^*G_3G_3
\end{equation}
This is nothing more that the complete supergravity action of Eq.~(\ref{sugra}).
 
This action is also  invariant under $\s_\bp$ and $\s_{pq}$, since it is equivalent to the one determined to be invariant under the complete untwisted BRST symmetry operator in~\cite{baubel}.
The transformations under all twisted supersymmetry generators of the fields are: 
\begin{equation}
\begin{array}{|c||c|c|c|}\hline
& \s & \s_{\bp} & \s_{pq} \\ \hline\hline
e_m & -\Psi_m & iJ_{m\bp}\Psi_0 & 0 \\
e_\bm & 0 & \Psi_{\bp\bm} & -2iJ_{\bm[p}\Psi_{q]} \\ \hline
\Psi_m & 0 & \; iJ_{\bp m}\left(\tfrac{1}{2}\o - A\right) + \o_{\bp m} \; & 0 \\
\Psi_0 & \tfrac{1}{2}\o - A & 0 & -\o_{pq} \\
\Psi_{\bm\bn} & \o_{\bm\bn} & 0 & 2J_{\bm [p|}J_{\bn |q]}\left(\tfrac{1}{2}\o + A\right) \\ \hline
\; \o_{mn} \; & X_{[m,n]} & -\tfrac{i}{2}J_{m\bp}\left(\rho_{\bq n,0}e_q + \rho_{qn,0}e_\bq\right) - \tfrac{1}{2}G_{mn\bp}\Psi_0 & 0 \\
\; \o_{\bm\bn} \; & 0 & \; -\tfrac{1}{2}\left(\rho_{\bq\bn,\bm\bp}e_q + \rho_{q\bn,\bm\bp}e_\bq + G_{\bm\bn q}\Psi_{\bp\bq}\right) & -iJ_{\bm[p}X_{\bn,|q]} \\
\; \o_{m\bn}\; & 2X_{m,\bn} & \; \tfrac{i}{2}J_{m\bp}\left(\rho_{\bq\bn,0}e_q + \rho_{q\bn,0}e_\bq\right) + \tfrac{1}{2}\left(G_{m\bp\bn}\Psi_0 - G_{mq\bn}\Psi_{\bp\bq}\right) \; & 0 \\ \hline
A & X_{m,\bm} & \; \tfrac{1}{2}\left(\rho_{\bq\bp,0}e_q + \rho_{q\bp,0}e_\bq - G_{m\bp\bm}\Psi_0 + G_{mq\bm}\Psi_{\bp\bq}\right) \; & X_{p,q} \\ \hline 
B_2 & -\Psi_m e_\bm & -\Psi_0e_\bp - \Psi_{\bp\bm}e_m & -2\Psi_{[p}e_{q]} \\ \hline
\end{array}
\nn
\end{equation}
with the twisted $X$ spinor in~(\ref{X}) defined as 
\begin{equation}
X_{m,n} = -\frac{1}{2}\left(\rho_{pn,m}e_\bp + \rho_{\bp n,m}e_p + G_{mn\bp}\Psi_p\right)
\end{equation}
Since these transformations are obtained directly from the Bianchi identities and the modified horizontality conditions for  field strengths, the three anticommutation relations~(\ref{anticomm}) hold true~\footnote{The explicit verification is non trivial, since it relies on the expression of the spin-connection, expressed as a solution of Eq.~(\ref{torsion_non_zero}).}.

\section {Matter and vector multiplets coupled to supergravity}

In this section, we will compute  both the scalar and vector symmetries acting on the matter fields, so we will retain $(\chi_0,\chi_p) \neq 0$ when we expand the curvature equations in ghost number. The invariant actions for both multiplets  can be expressed as  $\delta$ exact terms, in a way that generalizes the flat  space case  \cite{lbN}.

\subsection{The Wess--Zumino  multiplet}  

The Wess--Zumino matter multiplet is $(P,\sigma,H)$  where $P$ is   a complex scalar field, $\sigma $ a Majorana spinor (higgsino) and $H$ a complex  auxiliary field,  twisted into  $(\phi, \bar{\phi}, \sigma_0, \sigma_\bm, \sigma_{mn}, B_{\bm\bn},B_{mn})$. The various field strengths are  
\begin{eqnarray}
\hat{P} &=& \hat{D}\phi + \hat{\Psi}_m\sigma_\bm \CR
\hat{\bar{P}} &=& \hat{D}\bar{\phi} - \hat{\Psi}_0\sigma_0 - \hat{\Psi}_{\bm\bn}\sigma_{mn} \CR
\hat{\Sigma}_0 &=& \hat{D}\sigma_0 + B_{mn}\hat{\Psi}_{\bm\bn} \CR
\hat{\Sigma}_\bm &=& \hat{D}\sigma_\bm - B_{\bm\bn}\hat{\Psi}_n \\
\hat{\Sigma}_{mn} &=& \hat{D}\sigma_{mn} + B_{mn}\hat{\Psi}_0 \CR
\hat{H}_{mn} &=& \hat{D}B_{mn} \CR
\hat{H}_{\bm\bn} &=& \hat{D}B_{\bm\bn} \nn
\end{eqnarray}
with  the   covariant derivative $D$ explicitly defined as
\begin{eqnarray}
\hat{D}\phi &=& \hat{d}\phi + w\hat{A}\phi \CR
\hat{D}\bar{\phi} &=& \hat{d}\bar{\phi} - w\hat{A}\bar{\phi} \CR
\hat{D}\sigma_0 &=& \hat{d}\sigma_0 + \bigl(\tfrac{1}{2}\,\hat{\o} - w'\hat{A}\bigr)\sigma_0 + \hat{\o}_{\bm\bn}\sigma_{mn} \CR
\hat{D}\sigma_\bm &=& \hat{d}\sigma_\bm + \bigl(\tfrac{1}{2}\,\hat{\o} + w'\hat{A}\bigr)\sigma_\bm - \hat{\o}_{\bm n}\sigma_\bn \\
\hat{D}\sigma_{mn} &=& \hat{d}\sigma_{mn} - \bigl(\tfrac{1}{2}\,\hat{\o} + w'\hat{A} \bigr)\sigma_{mn} - \hat{\o}_{mn}\sigma_0 \CR
\hat{D}B_{mn} &=& \hat{d}B_{mn} - w''\hat{A}B_{mn} \CR
\hat{D}B_{\bm\bn} &=& \hat{d}B_{\bm\bn} + w''\hat{A}B_{\bm\bn} \nn
\end{eqnarray}
To have Bianchi identities, one must have  \(w' = w+1\) and \(w'' = w+2\). One obtains:
\begin{eqnarray}
\hat{D}\hat{P} &=& w\hat{F}\phi + \hat{\rho}_m\sigma_\bm - \hat{\Psi}_m\hat{\Sigma}_\bm \CR
\hat{D}\hat{\bar{P}} &=& -w\hat{F}\bar{\phi} - \hat{\rho}_0\sigma_0 - \hat{\rho}_{\bm\bn}\sigma_{mn} + \hat{\Psi}_0\hat{\Sigma}_0 + \hat{\Psi}_{\bm\bn}\hat{\Sigma}_{mn} \CR
\hat{D}\hat{\Sigma}_0 &=& \Bigl(\frac{1}{2}\hat{R} - (w+1)\hat{F} \Bigr)\sigma_0 + \hat{R}_{\bm\bn}\sigma_{mn} + \hat{H}_{mn}\hat{\Psi}_{\bm\bn} + B_{mn}\hat{\rho}_{\bm\bn} \CR
\hat{D}\hat{\Sigma}_\bm &=& \Bigl(\frac{1}{2}\hat{R} + (w+1)\hat{F}\Bigr)\sigma_\bm - \hat{R}_{\bm n}\sigma_\bn - \hat{H}_{\bm\bn}\hat{\Psi}_n - B_{\bm\bn}\hat{\rho}_n \\
\hat{D}\hat{\Sigma}_{mn} &=& \Bigl(-\frac{1}{2}\hat{R} - (w+1)\hat{F} \Bigr)\sigma_{mn} - \hat{R}_{mn}\sigma_0 + \hat{H}_{mn}\hat{\Psi}_0 + B_{mn}\hat{\rho}_0 \CR
\hat{D}\hat{H}_{mn} &=& -(w+2)\hat{F}B_{mn} \CR
\hat{D}\hat{H}_{\bm\bn} &=& (w+2)\hat{F}B_{\bm\bn} \nn
\end{eqnarray}
The distorted horizontality conditions that are compatible with the Bianchi identities and warrant off-shell closure,  are the following:
\begin{eqnarray}
\hat{P} &=& P \CR
\hat{\bar{P}} &=& \bar{P} \CR
\hat{\Sigma}_0 &=& \Sigma_0 + \hat{\Psi}_p(\bar{P}_\bp - \tfrac{w}{2}G_{m\bp\bm}\bar{\phi}) \CR
\hat{\Sigma}_\bm &=& \Sigma_\bm - \hat{\Psi}_0(P_\bm + \tfrac{w}{2}G_{p\bm\bp}\phi) \\
\hat{\Sigma}_{mn} &=& \Sigma_{mn} + \hat{\Psi}_{[m}(\bar{P}_{n]} - \tfrac{w}{2}G_{n]q\bq}\bar{\phi}) \CR
\hat{H}_{mn} &=& H_{mn} + \hat{\Psi}_p(\Sigma_{\bp,mn} + S_{\bp,mn}) - i\hat{\Psi}_pJ_{\bp[m}(\Sigma_{n],0} + S_{n],0}) \CR
\hat{H}_{\bm\bn} &=& H_{\bm\bn} - 2\hat{\Psi}_0(\Sigma_{[\bm,\bn]} + S_{[\bm,\bn]}) \nn
\end{eqnarray}
where
\begin{eqnarray}
S_{\bp,mn} - iJ_{\bp[m}S_{n],0} = &&2iwG_{q\br\bq}J_{\bp[m}\Psi_{n],r}\bar{\phi} - i\tfrac{w}{2}J_{\bp[m}G_{n]q\bp}\Psi_{\bq,p}\bar{\phi} + i\tfrac{w}{2}J_{\bp[m}G_{n]q\bq}\sigma_0 \CR
&&-\tfrac{w+2}{2}G_{\bq q\bp}\sigma_{mn} + G_{m\bp n}\sigma_0 \CR
S_{[\bm,\bn]} = &&(P_{[\bm} - \tfrac{w}{2}G_{q\bq[\bm})\Psi_{\bn],0} + (P_q - \tfrac{w}{2}G_{rq\br})\Psi_{[\bm,\bn]\bq} - G_{q\bq[\bm}\sigma_{\bn]} \CR
&&- G_{\bm\bn p}\sigma_\bp + \tfrac{w}{2}\rho_{\bm\bn,0}\phi - \tfrac{w}{2}\rho_{q\bq,\bm\bn}\phi \nn
\end{eqnarray}
The ghost number 1 parts of these equations give the scalar and vector transformations of the fields
\begin{align}
\label{matter_transfo}
\s\phi &= 0 & \s_\bp\phi &= -\sigma_\bp \CR
\s\bar{\phi} &= \sigma_0 & \s_\bp\bar{\phi} &= 0 \CR
\s\sigma_0 &= 0 & \s_\bp\sigma_0 &= \bar{P}_\bp - \tfrac{w}{2}G_{m\bp\bm}\bar{\phi} \CR
\s\sigma_\bm &= -P_\bm - \tfrac{w}{2}G_{p\bm\bp}\phi & \s_\bp\sigma_\bm &= B_{\bp\bm} \\
\s\sigma_{mn} &= -B_{mn} & \s_\bp\sigma_{mn} &= i(\bar{P}_{[m|} - \tfrac{w}{2}G_{q\bq[m|}\bar{\phi})J_{n]\bp} \CR
\s B_{mn} &= 0 & \s_\bp B_{mn} &= (\Sigma_{\bp,mn} + S_{\bp,mn}) - iJ_{\bp[m}(\Sigma_{n],0} + S_{n],0}) \CR
\s B_{\bm\bn} &= -2(\Sigma_{[\bm,\bn]} + S_{[\bm,\bn]}) & \s_\bp B_{\bm\bn} &= 0 \nn
\end{align}
The anticommutation relations~(\ref{anticomm}) can be explicitly verified on all fields, in a much   easier way than for the supergravity multiplet (see the Appendix D).

\subsection{The vector multiplet}

The twisted   vector multiplet is $(B,\xi_m,\xi_{\bm\bn},\xi_0,h)$, with $B$ a $U(1)$ gauge field, $(\xi_m,\xi_{\bm\bn},\xi_0)$ its twisted Majorana supersymmetric partner and $h$ a real auxiliary field.  The field strengths are 
\begin{eqnarray}
\hat{\mathcal{F}} &=& \hat{d}B - (\hat{\Psi}_0\xi_m + \hat{\Psi}_m\xi_0)e_\bm - (\hat{\Psi}_p\xi_{\bm\bp} + \hat{\Psi}_{\bm\bp}\xi_p)e_m \CR
\hat{\Xi}_0 &=& \hat{D}\xi_0 - h\hat{\Psi}_0 \CR
\hat{\Xi}_m &=& \hat{D}\xi_m + h\hat{\Psi}_m \\
\hat{\Xi}_{\bm\bn} &=& \hat{D}\xi_{\bm\bn} - h\hat{\Psi}_{\bm\bn} \CR
\hat{\mathcal{H}} &=& \hat{d}h \nn
\end{eqnarray}
where $\hat D$ is given by $\hat{D}\xi_0 = \hat{d}\xi_0 - \tfrac{1}{2}\hat{\o}\xi_0 + \hat{A}\xi_0 + \hat{\o}_{mn}\xi_{\bm\bn}$, etc. The Bianchi identities for these field strengths are
\begin{eqnarray}
\hat{d}\hat{\mathcal{F}} &=& \Bigl(\hat{\Psi}_0\hat{\Xi}_m + \hat{\Psi}_m\hat{\Xi}_0 - \hat{\rho}_0\xi_m - \hat{\rho}_m\xi_0\Bigr)\hat{e}_\bm + \Bigl(\hat{\Psi}_p\hat{\Xi}_{\bm\bp} + \hat{\Psi}_{\bm\bp}\hat{\Xi}_p - \hat{\rho}_p\xi_{\bm\bp} - \hat{\rho}_{\bm\bp}\xi_p\Bigr)\hat{e}_m \CR
&&+ (\hat{\Psi}_m\xi_0 + \hat{\Psi}_0\xi_m)\hat{T}_\bm + (\hat{\Psi}_{\bm\bp}\xi_p + \hat{\Psi}_p\xi_{\bm\bp})\hat{T}_m \CR
\hat{D}\hat{\Xi}_0 &=& \bigl(-\tfrac{1}{2}\hat{R} + \hat{F}\bigr)\xi_0 + \hat{R}_{mn}\xi_{\bm\bn} - \hat{\mathcal{H}}\hat{\Psi}_0 - h\hat{\rho}_0 \\
\hat{D}\hat{\Xi}_m &=& -\bigl(\tfrac{1}{2}\hat{R} + \hat{F} \bigr)\xi_m - \hat{R}_{\bp m}\xi_p + \hat{\mathcal{H}}\hat{\Psi}_m + h\hat{\rho}_m \CR
\hat{D}\hat{\Xi}_{\bm\bn} &=& \bigl(\tfrac{1}{2}\hat{R} + \hat{F} \bigr)\xi_{\bm\bn} - \hat{R}_{\bm\bn}\xi_0 - \hat{\mathcal{H}}\hat{\Psi}_{\bm\bn} - h\hat{\rho}_{\bm\bn} \CR
\hat{d}\hat{\mathcal{H}} &=& 0 \nn
\end{eqnarray}
The supersymmetry is defined by the constraints
\begin{eqnarray}
\hat{\mathcal{F}} &=& \mathcal{F} \CR
\hat{\Xi}_0 &=& \Xi_0 + \mathcal{F}_{mn}\hat{\Psi}_{\bm\bn} \CR
\hat{\Xi}_m &=& \Xi_m - \mathcal{F}_{\bp m}\hat{\Psi}_p \\
\hat{\Xi}_{\bm\bn} &=& \Xi_{\bm\bn} + \mathcal{F}_{\bm\bn}\hat{\Psi}_0 \CR
\hat{\mathcal{H}} &=& \mathcal{H} + \hat{\Psi}_p(\Xi_{\bp,0} + G_{m\bp n}\xi_{\bm\bn}) \nn
\end{eqnarray}
which give
\begin{align}
\label{vector_transfo}
\s B &= \xi_me_\bm & \s_\bp B &= \xi_0e_\bp + \xi_{\bm\bp}e_m \CR
\s\xi_0 &= h & \s_\bp\xi_0 &= 0 \CR
\s\xi_m &= 0 & \s_\bp\xi_m &= \mathcal{F}_{\bp m} - iJ_{\bp m}h \\
\s\xi_{\bm\bn} &= \mathcal{F}_{\bm\bn} & \s_\bp\xi_{\bm\bn} &= 0 \CR
\s h &= 0 & \s_\bp h &= \Xi_{\bp,0} + G_{m\bp n}\xi_{\bm\bn} \nn
\end{align}
The algebra closure relations~(\ref{anticomm}) are satisfied on all fields (see the Appendix D).

\section{Conclusion and outlook} 

We have shown that the supergravity action is essentially determined by its invariance under a single  scalar supersymmetry generator. This scalar generator is nilpotent and formally similar to a BRST operator. It is singled out from the multiplet of supersymmetry generators by a twist and is therefore quite analogous to the one encountered in the twisted super-Yang--Mills theory in four dimensions.  
The supergravity action has  parts  which are independently invariant under this scalar generator and induce an interesting decomposition of both the Einstein and Rarita--Schwinger actions in twisted form. 

In the twisted form, there is also a vector supersymmetry generator $\delta_{\bar p }$. Its anticommutation with the scalar generator gives rise to translations, but with additional field dependent gauge transformations. These commutation relations are best related to the BRST transformations of the ghost fields, with a consistency derived from Bianchi identities.  Nevertheless, when the gravitino field vanishes, the fourth symmetry $\delta_{mn}$ can be safely ignored.  This additional symmetry does not add any new constraint to the action. 

There is an  underlying localization around gravitational instantons  that seems of interest in this construction.  A  twisted formulation of  the Wess--Zumino and vector multiplets coupled to the supergravity multiplet has also been obtained. 
Generalizations to higher dimensional supergravities could be of interest and a analogous twist could be used to split the Poincar\'{e} symmetry of, for example, \(d=10\) supergravity into smaller and (hopefully) simpler sectors.

\section*{Acknowledgments}
The work of V. R. is presently supported by the ERC Advanced Grant No. 246974, \textit{'Supersymmetry: a window to non-perturbative physics'}. 
 
\appendix
\numberwithin{equation}{section}
  
\section{The BSRT symmetry from horizontality conditions}
 
The supergravity transformations can be expressed as BRST transformations, in a way that merely generalizes  the Yang--Mills case (ghost unification, horizontality equations for the curvatures, etc.)~\cite{baubel}.  Call $s$ the BRST operator of the supergravity transformation, and its ghost $\xi$. The other ghosts are those of local SUSY ($\chi$), Lorentz symmetry ($\Omega$), the chiral \(U(1)\) symmetry ($c$)  and the 2-form gauge symmetry ($B^1_1$).  One gets the usual  transformation laws of classical fields by changing the   ghosts into local parameters, with the opposite statistics. Their  off-shell closure property is equivalent to the nilpotency of the graded differential operator $s$. The difficult part of the supergravity BRST symmetry is its dependence on the supersymmetry ghost $\chi$. The reparametrization invariance can be absorbed, by redefining $\hat s$ as $\hat s = s -\cal{L} _{\xi}$, with $s\xi^\mu = \xi^\nu\partial_\nu\xi^\mu + \frac{1}{2}\bar{\chi}\gamma^\mu \chi$. With this property, the off-shell closure relation $s^2=0$ is equivalent to $\hat s ^2 = \cal {L}_{\bar{\chi}\gamma^\mu\chi}$. Reparametrization invariance is decoupled by the operation $\textnormal{exp}(-i_{\xi})$, when classical and ghost fields are unified into graded sums, a property that was found for the study of gravitational anomalies but turns out to be very useful for the construction of supergravity BRST symmetries. For the $N=1,d=4$ supergravity in the new minimal scheme, the action of the operator $\hat s$ is as follows
\begin{eqnarray}
\sg e^a &=& -\Omega^{ab}e_b - i\bar{\chi}\gamma^a\lambda \CR 
\sg\lambda &=& -D\chi - \Omega^{ab}\gamma_{ab}\lambda - c\gamma^5\lambda \CR
\sg B_2 &=& -dB_{1}^1 - i\hat{\chi}\gamma^a\lambda e_a \\
\sg A &=& -dc - \tfrac{1}{2}i\bar{\chi}\gamma^5\gamma^aX_a \CR 
\sg \omega^{ab} &=& -(D\Omega)^{ab} - i\bar{\chi}\gamma^{[a}X^{b]} \nn
\end{eqnarray}
where the spinor $X_a$ is $X_a  =\rho_{ab}e^b -(\tfrac{1}{2}G_{abc}\gamma^{bc} + \tfrac{1}{12}\epsilon_{abcd}G^{bcd}\gamma^5)\lambda$. $X_a$ vanishes when one uses the equations of motion of the gravitino and of the (propagating) auxiliary fields. The property $s^2=0$, equivalent to $\hat s ^2= \cal{L}_{\bar{\chi}\gamma^\mu\chi}$ is warranted by the ghost transformation laws~\cite{baubel}. At the root of these equations, there  is a unification between classical fields and ghosts~\cite{baubel}, which is analogous to the one that occurs when analyzing anomalies by descent equations. In fact, everything boils down to computing constraints on the curvatures, which satisfy the following Bianchi identities:
\begin{eqnarray} 
\label{conssugra}
\hat T^a &\equiv& \hat d  e^a + (\omega+\Omega)^{ab}e_b + \frac{i}{2}(\bar\la +\bar\chi)\gamma^a (\la +\chi) = -\frac{1}{2} G^a_{bc}e^b e^c \CR
\hat \rho &\equiv& \hat d (\la +\chi) + (\omega+\Omega+A+c)(\la +\chi) = \frac{1}{2}\rho_{ab} e^ae^b  \CR 
\hat G_3 &\equiv& \hat d (B_2+B^1_1+B_0^2) + \frac{i}{2}  (\bar\la +\bar\chi)\gamma^a (\la +\chi)  e^a = \frac{1}{6} G_{abc}e^a e^b e^c \\
\hat R^{ab} &\equiv& \hat d (\omega+\Omega) + (\omega+\Omega)^2 = R^{ab} - i\bar\chi \gamma^{[a}  X^{b]} - \frac{i}{4}\bar\chi\gamma^{c}\chi G^{ab}_c \CR 
\hat F &\equiv& \hat d(A+c) = F - \frac{i}{2}\bar\chi\gamma^5\gamma^{a}X_a - \frac{i}{24}\bar\chi\gamma^{a}\chi\epsilon_{abcd}G^{bcd} \nn
\end{eqnarray}
By expansion at ghost number one, one finds the transformation laws in Eq.~(\ref{brs}) and at ghost number two, one finds those of the ghosts:
\begin{eqnarray}
\sg\chi &=& -i_\phi \la - \Omega\chi - c\chi \CR
\sg c &=& -i_\phi A - \tfrac{i}{24}\bar{\chi}\gamma^a\chi\epsilon_{abcd}G^{bcd} \CR 
\sg B_1^1 &=& -i_\phi B - dB_0^2 - \tfrac{i}{2}\bar{\chi}\gamma^a\chi e_a \\
\sg B^2 &=& -i_\phi B_1^1 \CR
\sg \Omega^{ab} &=& -i_\phi\omega^{ab} -\tfrac{1}{2}\left[\Omega,\Omega\right]^{ab} - \tfrac{i}{2}\bar{\chi}\gamma^c\chi G_c^{ab} \nn
\end{eqnarray}

\section {Tensor and chirality conventions}
\label{conventions}

The normalization of the completely antisymmetric four-index symbol with tangent space indices is\begin{equation}\epsilon_{0123} = 1
\end{equation}
Once twisted, this is taken to be\begin{equation}\epsilon_{1\bar{1}2\bar{2}} = 1
\end{equation}
The dual of an antisymmetric Lorentz tensor is\begin{equation}\tilde{F}_{ab} = \frac{1}{2}\epsilon_{abcd}F^{cd}
\end{equation}
The selfdual and antiselfdual parts of $F_{ab}$ are\begin{equation}F_{ab}^{\pm} = \frac{1}{2}(F_{ab} \pm \tilde{F}_{ab})
\end{equation}
We take $\gamma_5$ such that $(\gamma_5)^2 = -1$ and define the chiral projections
\begin{eqnarray}
\lambda^{\pm} &=& \frac{1 \pm i\gamma_5}{2}\lambda \CR
\bar{\lambda}^{\pm} &=& \bar{\lambda}\frac{1 \pm i\gamma_5}{2}
\end{eqnarray}
in order to have $\lambda = \lambda^+ + \lambda^-$ and $\bar{\lambda} = \bar{\lambda}^+ + \bar{\lambda}^-$. Then, we have the useful identity
\begin{eqnarray}
\bar{\lambda}^+ \gamma^a \lambda^+ = \bar{\lambda}^+ \gamma_5 \gamma^a \gamma_5 \lambda^+ = -i\bar{\lambda}^+ \gamma^a (-i)\lambda^+ = -\bar{\lambda}^+ \gamma^a \lambda^+ = 0
\end{eqnarray}
and similarly $\bar{\lambda}^- \gamma^a \lambda^- = 0$. Finally, once in twisted form, the chiral projections of spinor separate its various components according to
\begin{eqnarray}
\lambda^+ &\sim& (0,\Psi_p,0) \CR
\lambda^- &\sim& (\Psi_0,0,\Psi_{\bm\bn}) \nn
\end{eqnarray} 

\section{The action of $\gamma$ matrices on twisted spinors}
\label{twisted_gamma}

The action of a $\gamma$ matrix on a twisted spinor with components $(\Psi_0, \Psi_m, \Psi_{\bm\bn})$ is defined as follows\begin{equation}\begin{array}{|c||c|c|c|}
\hline
& 0 & \bp & pq \\
\hline
\: \gamma_m\Psi \: & \: i\Psi_m \: & \: -J_{m\bp}\Psi_0 \: & \: 0 \: \\
\: \gamma_\bm\Psi \: & \: 0 \: & \: i\Psi_{\bm\bp} \: & \: 2J_{\bm[p}\Psi_{q]} \: \\
\hline
\end{array}
\end{equation}
Similarly, the action of a $\gamma$ matrix on a twisted spinor with components $(\sigma_0, \sigma_\bm, \sigma_{mn})$, as the one appearing in the Wess--Zumino multiplet, is
\begin{equation}
\begin{array}{|c||c|c|c|}
\hline
& 0 & p & \bp\bq \\
\hline
\: \gamma_m\sigma \: & \: 0 \: & \: i\sigma_{mp} \: & \: -J_{m[\bp}\sigma_{\bq]} \: \\
\: \gamma_\bm\sigma \: & \: i\sigma_\bm \: & \: -J_{p\bm}\sigma_0 \: & \: 0 \: \\
\hline
\end{array}
\end{equation}
These conventions allow us to retrieve the Clifford algebra for the twisted $\gamma$ matrices
\begin{eqnarray}
\{\gamma_m,\gamma_n\} &=& 0 \CR
\{\gamma_\bm,\gamma_\bn\} &=& 0 \CR
\{\gamma_m,\gamma_\bn\} &=& -iJ_{m\bn} \equiv g_{m\bn} \nn
\end{eqnarray} 
We also define the $\gamma_{ab}$ matrices in twisted form as
\begin{eqnarray}
\gamma_{m\bn} &=& \gamma_m\gamma_\bn - \gamma_\bn\gamma_m \CR
\gamma_{mn} &=& \gamma_m\gamma_n - \gamma_n\gamma_m \CR
\gamma_{\bm\bn} &=& \gamma_\bm\gamma_\bn - \gamma_\bn\gamma_\bm \nn
\end{eqnarray}
which act on the two kinds of twisted spinors according to the following tables
\begin{equation}
\begin{array}{|c||c|c|c|}
\hline
& 0 & p & \bp\bq \\
\hline
\: \gamma_{mn}\Psi \: & \: 0 \: & \: 0 \: & \: -2J_{m[\bp}J_{\bq]n}\Psi_0 \: \\
\: \gamma_{\bm\bn}\Psi \: & 2\Psi_{\bm\bn} & 0 & 0 \\
\: \gamma_{m\bn}\Psi \: & \: iJ_{m\bn}\Psi_0 \: & \: 2iJ_{p\bn}\Psi_m- iJ_{m\bn}\Psi_p \: & \: -iJ_{m\bn}\Psi_{\bp\bq} \: \\
\hline
\end{array}
\end{equation}
\begin{equation}\begin{array}{|c||c|c|c|}
\hline
& 0 & \bp & pq \\
\hline
\: \gamma_{mn}\sigma \: & \: 2\sigma_{mn} \: & \: 0 \: & \: 0 \: \\
\: \gamma_{\bm\bn}\sigma \: & 0 & 0 & \: -2J_{\bm[p}J_{q]\bn}\sigma_0 \: \\
\: \gamma_{m\bn}\sigma \: & \: -iJ_{m\bn}\sigma_0 \: & \: -\tfrac{3i}{2}J_{m\bp}\sigma_\bn + \tfrac{i}{2}J_{m\bn}\sigma_\bp \: & \: iJ_{m\bn}\sigma_{pq} \: \\
\hline
\end{array}
\end{equation}

\section{Algebra closure on the  fields of matter and vector multiplets}

In this appendix, we give some examples of the anticommutation relations~(\ref{anticomm}) on some matter fields of the Wess--Zumino and vector multiplets.

Starting with the $\phi$ and $\bar{\phi}$ fields of the Wess--Zumino multiplet, one needs their transformation laws under the pseudo-scalar symmetry in order to check~(\ref{anticomm}). These are obtained in the same way as the scalar and vector symmetry transformation laws, \textit{i.e.} by isolating the part of ghost number 1 in the horizontality conditions on $\hat{P}=P$ and $\hat{\bar{P}}=\bar{P}$ and keeping $\chi_{mn} \neq 0$. This yields
\be
\label{tensor_symm_phi}
\delta_{mn}\phi = 0 \quad \quad \quad \textnormal{and} \quad \quad \quad \delta_{mn}\bar{\phi} = \sigma_{mn}
\ee

The tranformation laws in~(\ref{matter_transfo}) allow us to compute straightforwardly 
\bea
\delta^2\phi &=& 0 \CR
\{\delta_{\bp},\delta_{\bq}\}\phi &=& -\left(B_{\bp\bq} + B_{\bq\bp}\right) = 0 \\
\{\delta,\delta_\bp\}\phi &=& \left(P_\bp + \frac{w}{2}G_{m\bp\bm}\phi\right) \CR
&=& \partial_\bp\phi + \left(wA_\bp + \frac{w}{2}G_{m\bp\bm}\right)\phi + \Psi_{\bp,m}\sigma_\bm \CR
&=& \partial_\bp\phi + \delta^{\textnormal{gauge}}(A,G)\phi - \displaystyle\sum\limits_{a= 0,m,\bm\bn}\Psi_{\bp,a}\delta_\ba\phi \nn
\eea
where the last equality is a consequence of~(\ref{matter_transfo}) and~(\ref{tensor_symm_phi}).

Similarly, on $\bar{\phi}$:
\bea
\delta^2\bar{\phi} &=& \delta\sigma_0 = 0 \CR
\{\delta_{\bp},\delta_{\bq}\}\bar{\phi} &=& 0 \\
\{\delta,\delta_\bp\}\bar{\phi} &=& \left(\bar{P}_\bp - \frac{w}{2}G_{m\bp\bm}\bar{\phi}\right) \CR
&=& \partial_\bp\bar{\phi} - \left(wA_\bp + \frac{w}{2}G_{m\bp\bm}\right)\bar{\phi} - \left(\Psi_{\bp,0}\sigma_0 + \Psi_{\bp,\bm\bn}\sigma_{mn}\right) \CR
&=& \partial_\bp\bar{\phi} + \delta^{\textnormal{gauge}}(A,G)\bar{\phi} - \displaystyle\sum\limits_{a= 0,m,\bm\bn}\Psi_{\bp,a}\delta_\ba\bar{\phi} \nn
\eea
using again~(\ref{matter_transfo}) and~(\ref{tensor_symm_phi}) for the last equality.\\

Turning to the $B$ field of the vector multiplet, the horizontality condition on its field strength $\hat{\mathcal{F}} = \mathcal{F}$ allows us to compute
\be
\label{tensor_symm_B}
\delta_{mn}B = \xi_n e_m
\ee
and the transformation laws~(\ref{vector_transfo}) of the vector multiplet fields yield:
\bea
\delta^2B &=& \delta(\xi_m e_\bm) = 0 \CR
\{\delta_{\bp},\delta_{\bq}\}B &=& \xi_0\Psi_{\bp\bq} + \xi_{\bp\bq}\Psi_0 + \xi_0\Psi_{\bq\bp} + \xi_{\bq\bp}\Psi_0 = 0 \\
\{\delta,\delta_\bp\}B &=& he_\bp + \mathcal{F}_{\bm\bp}e_m - \xi_{\bm\bp}\Psi_m + \mathcal{F}_{\bp m}e_\bm - iJ_{\bp m}he_\bm + \xi_m\Psi_{\bp\bm} \CR
&=& \mathcal{F}_{\bp m}e_\bm - \mathcal{F}_{\bp\bm}e_m - \xi_{\bm\bp}\Psi_m + \xi_m\Psi_{\bp\bm} \CR
&=& \partial_\bp B - \left(\Psi_{\bp,0}\xi_m + \Psi_{\bp,m}\xi_0\right)e_\bm - \left(\Psi_{\bp,q}\xi_{\bm\bq} + \Psi_{\bp,\bm\bq}\xi_q\right)e_m \CR
&=& \partial_\bp B - \displaystyle\sum\limits_{a= 0,m,\bm\bn}\Psi_{\bp,a}\delta_\ba B \nn
\eea
where, for the last equality, we've used the $B$ transformations given by~(\ref{vector_transfo}) and~(\ref{tensor_symm_B}).

\end{document}